\documentclass[sigconf]{acmart}

\usepackage[utf8]{inputenc}
\usepackage[T1]{fontenc}

\usepackage[
    nameinlink  %
]{cleveref}
\Crefname{equation}{Eq.}{Eqs.}
\Crefname{figure}{Fig.}{Figs.}
\Crefname{table}{Tab.}{Tabs.}
\Crefname{appendix}{App.}{Apps.}
\Crefname{section}{Sec.}{Secs.}

\usepackage{colortbl}
\usepackage{multirow}
\usepackage{tabularx}
\usepackage{siunitx}
\DeclareSIUnit[per-mode=symbol]\bps{\bit \per \second}
\DeclareSIUnit[per-mode=symbol]\kbps{\kilo\bps}
\DeclareSIUnit[per-mode=symbol]\Mbps{\mega\bps}
\DeclareSIUnit[per-mode=symbol]\Gbps{\giga\bps}
\DeclareSIUnit{\nothing}{\relax}

\newcommand{\mn}{\mega\nothing}
\newcommand{\kn}{\kilo\nothing}

\newcommand{\sk}[1]{\SI{#1}{\kn}}
\newcommand{\sm}[1]{\SI{#1}{\mn}}

\newcommand{\sperc}[1]{\SI{#1}{\percent}}

\usepackage{xspace}
\usepackage{tikz}
\usetikzlibrary{fit, positioning, shapes}
\usetikzlibrary{arrows, decorations.markings}
\usetikzlibrary{patterns}

\usepackage{enumitem}
\usepackage{balance}

\usepackage{acro}
\DeclareAcronym{apd}{
    short = APD,
    long = Aliased Prefix Detection
}
\DeclareAcronym{mptcp}{
    short = MPTCP,
    long = Multipath TCP
}
\DeclareAcronym{gfw}{
    short = GFW,
    long = Great Firewall of China
}
\DeclareAcronym{tga}{
    short = TGA,
    short-plural-form = TGAs,
    long = Target Generation Algorithm,
    long-plural-form = Target Generation Algorithms
}

\newcommand*{\eg}{\emph{e.g.,}\@\xspace}

\newcommand*{\ie}{\emph{i.e.,}\@\xspace}

\newcommand{\parx}[1]{\vspace{0.5ex}\noindent{\textbf{#1}}}

\newcommand*{\hl}{IPv6 Hitlist Service\xspace}

\setcopyright{none}
\renewcommand\footnotetextcopyrightpermission[1]{}

\usepackage{fancyhdr}
\newcommand{\ccrfooter}{%
   \fancyhf{}%
   \renewcommand{\headrulewidth}{0pt}%
   \renewcommand{\footrulewidth}{0pt}%
   \fancyfoot[L]{\footnotesize ACM SIGCOMM Computer Communication Review}%
   \fancyfoot[R]{\footnotesize Volume 56 Issue 3, October 2026}%
}
\AtBeginDocument{%
   \fancypagestyle{firstpagestyle}{\ccrfooter}%
   \fancypagestyle{standardpagestyle}{\ccrfooter}%
   \pagestyle{standardpagestyle}%
}

\ccsdesc[500]{Networks~Network measurement}
\keywords{IPv6, Hitlist, Internet scanning}

\usepackage[absolute,showboxes]{textpos}

\begin{document}

\settopmatter{printfolios=true}
\setlength{\TPHorizModule}{\paperwidth}
\setlength{\TPVertModule}{\paperheight}
\TPMargin{5pt}
\begin{textblock}{0.8}(0.1,0.02)
     \noindent
     \footnotesize
     If you cite this paper, please use the SIGCOMM CCR reference:
     Oliver Gasser, Lion Steger, Patrick Sattler, Johannes Zirngibl. 2026.
     IPv6 Hitlist Service: Lessons Learned From 10 Years of Operation.
     In \textit{ACM SIGCOMM Computer Communication Review, Volume 56, Issue 3, October 2026.}
     ACM, New York, NY, USA, 9 pages.
\end{textblock}
 
\title[IPv6 Hitlist Service: Lessons Learned From 10 Years of Operation]{IPv6 Hitlist Service:\\ Lessons Learned From 10 Years of Operation}

\author{Oliver Gasser}
\affiliation{
	\institution{IPinfo}
    \country{}
}
\email{oliver@ipinfo.io}

\author{Lion Steger}
\affiliation{
	\institution{Technical University of Munich}
    \country{}
}
\email{stegerl@net.in.tum.de}

\author{Patrick Sattler}
\affiliation{
	\institution{BENOCS}
    \country{}
}
\email{psattler@benocs.com}

\author{Johannes Zirngibl}
\affiliation{
	\institution{Max Planck Institute for Informatics}
    \country{}
}
\email{johannes.zirngibl@mpi-inf.mpg.de}

\begin{abstract}
    After becoming an Internet Draft more than 30 years ago, IPv6 has seen an increase in deployment and use in the past years.
    As measurements in the IPv6 Internet require new approaches due to the vastly larger address space, hitlists have come along as one possible source for finding IPv6 targets.
    One of the most prominent hitlists is provided by the \hl.

    In this paper, we share insights from 10 years of operations of the \hl:
    We show different evolutions of the service, highlighting important changes along the way.
    To better understand the representativeness of the hitlist, we perform a coverage analysis using real-world traffic data from a major central European ISP and Tier-1 network, finding that at least one address is known to the IPv6 Hitlist Service for \sperc{87.1} of ASes and \sperc{56.5} of /48 prefixes originating IPv6 traffic.
    We also share results from a conducted user survey and analyze users accessing the \hl, finding diverse use cases and access patterns across time (e.g., one-off vs. continuous downloads) and available data (e.g., all vs. responsive addresses).
    Finally, we provide best practice recommendations when working with the hitlist and share lessons learned during its 10-year operation.
\end{abstract}
 
\maketitle

\section{Introduction}

\begin{figure}
    \centering
    \includegraphics[width=\linewidth]{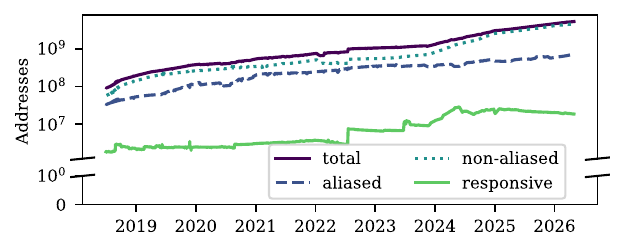}
    \vspace{-2em}
    \caption{Temporal development of the \hl.}
    \label{fig:hl-over-time}
    \vspace{-2em}
\end{figure}

In March 2026, IPv6 user traffic to Google services for the first time surpassed IPv4 traffic~\cite{googlev6growth}.
This milestone shows that IPv6---more than 30 years after becoming an Internet Draft~\cite{rfc1883}---has finally surpassed this important threshold.
An increase in IPv6 traffic is not unique to Google, but has been observed by other companies and organizations as well~\cite{cloudflarev6growth, apnicv6growth}. %
The many architectural changes introduced by IPv6 compared to its predecessor also affect the Internet measurement community.
One of these important changes is the address space, which in IPv6 is $10^{29}$ times larger compared to IPv4.
This means that while in IPv4, brute-force scanning of the complete address space is possible within hours~\cite{zmap,masscan}, it is not feasible in IPv6.
Therefore, IPv6 measurements usually use techniques such as hitlists~\cite{gasser2016scanning,gasser2018clusters,zirngibl2022rustyclusters,Steger2023,song2020towards,rye2023ipv6}, Target Generation Algorithms (TGAs)~\cite{6Sense, 6Scan, 6Tree,AddrMiner}, or similar approaches to perform smart target selection.

One of the most prominent examples of this is the \hl~\cite{v6hl}, which was first created back in 2016.
The initial IPv6 hitlist TMA 2016 paper~\cite{gasser2016scanning} focused on the analysis of potential hitlist sources and their differences and its service was run mostly internally, providing only partial access to IPv6 addresses upon request~\cite{v6hlorig}.
The follow-up IMC 2018 paper~\cite{gasser2018clusters} had a public IPv6 hitlist~\cite{v6hl}, which could be directly used by other researchers, as one of its main contributions.
The ongoing service was actively maintained and extended in the following years~\cite{zirngibl2022rustyclusters,Steger2023}.
As can be seen in \Cref{fig:hl-over-time}, the number of addresses has seen a large increase over time: 60x total addresses and 10x responsive addresses.
Moreover, since 2018, the \hl has seen more than 150 registered users and an even larger number of people downloading the publicly available version.
Furthermore, the hitlist has been instrumental in many IPv6 measurement research works and has been cited by more than 240 publications as of May 2026.
The majority of these publications are from the Internet measurement community (22x ACM IMC, 13x PAM, 3x TMA), but also the broader networking community (10x IEEE INFOCOM, 7x IEEE/ACM TON, 6x ACM CoNEXT) and the security community (3x USENIX Security, 2x IEEE S\&P, 2x IEEE EuroS\&P) are making use of the hitlist.

\noindent{}\textit{In this paper, we share insights on 10 years of operations of the \hl.
Specifically, we make the following main contributions:}
\begin{itemize}[nosep, leftmargin=*]
    \item \textbf{Changes Over Time:} We highlight different changes affecting the hitlist over time, such as additional data sources and changing characteristics of IPv6 networks (see \Cref{sec:technical}).
    \item \textbf{Coverage:} We compare the coverage of the hitlist to real-world traffic data from a major central European ISP and Tier-1 network. At least one address is known to the IPv6 Hitlist Service for \sperc{87} of ASes and \sperc{56} of /48 prefixes originating IPv6 traffic (see \Cref{sec:coverage}).
    \item \textbf{User Survey:} We report on results from a conducted user survey which shows different use cases of the hitlist data and underlines its importance for IPv6 measurement research (see \Cref{sec:survey}).
    \item \textbf{User Behavior:} We share insights on hitlist user behavior and show different access patterns (daily vs. one-off, all vs. responsive addresses) for un/registered users (see \Cref{sec:hitlistuse}).
    \item \textbf{Best Practices \& Lessons Learned:} We provide recommendations for best practices in hitlist usage and share lessons learned from building a successful measurement service (see \Cref{sec:bestpractices}).
\end{itemize}
\section{Technical Development}
\label{sec:technical}

In this section, we detail the architecture and evolution of the \hl.
This includes an analysis of the technical changes, modifications to the input sources and their respective impact on the data, and an automated analysis of trends in the data.

\parx{\hl Pipeline:}
The first step in the \hl pipeline is to combine different sources of addresses, which we further analyze in the following paragraphs, as well as the cumulative input from the last scans.
These addresses are filtered for duplicates, resulting in \num{5.24}~B addresses as of May 2026, which are published as the \emph{input} dataset.
After that, a blocklist filter is applied in order to enforce the opt-out mechanisms we implement for ethical scanning, followed by a filter for routable prefixes, during both of which only \sm{81} addresses are filtered out.
An overview can be seen in \Cref{fig:pipeline} in \Cref{app:pipeline}.

At this point, we apply our \ac{apd} method, where we sort all addresses into BGP prefixes and subnets of \texttt{/64} to \texttt{/120} in steps of 4 bits, considering any subnet which contains more than 100 addresses, except for \texttt{/64}, where all subnets are considered.
For all of these subnets, we generate 16 random addresses, one for each possible value \texttt{0--f} of the first nibble after the subnet part of the address.
We then scan all of these addresses with one ICMP and TCP/80 probe each and label the subnet as aliased if we receive a response for either probe for all 16 addresses in this or the previous two scans (see \cite{gasser2018clusters, zirngibl2022rustyclusters} for more details).
This step removes \sm{678} addresses.
Finally, any address that was unresponsive for more than 30 days is filtered, which avoids repeatedly scanning inactive addresses.
This step removes the majority (\num{4.02} B) of addresses, leaving \sm{458} addresses to be scanned.

We randomize the addresses to evenly distribute scanning traffic over networks and conduct traceroute measurements for all addresses.
We probe the addresses with different probe types, including ICMPv6 Echo Requests, TCP SYN packets on ports 80 and 443, and UDP packets with a QUIC initial packet for port 443 and a DNS query on port 53.
As a last step we send multipath TCP probes to ports 80 and 443~\cite{aschenbrenner2021from} and an SNMPv3 probe on UDP port 161~\cite{albakour2021snmp}.
All addresses responsive to at least one of these protocols (excluding traceroute) are published as \emph{responsive addresses} without registration, and the exact scan results for the different probes are published in the registration-first section.

\begin{table}
    \centering
    \footnotesize
    \caption{Summary of the most important changes to the \hl. Bold changes are visualized in subsequent plots.}
    \label{tab:changes}
    \begin{tabular}{l l}
        \toprule
        \textbf{Date} & \textbf{Summary} \\
        \midrule
        2021-03-22 & Addition of MPTCP scans \\ %
        2021-06-05 & Addition of UDP/161 scans \\ %
        2022-01-13 & \textbf{Removal of injected DNS responses} \\ %
        2023-03-24 & Change of DNS payload to avoid injections \\ %
        2024-03-16 & \textbf{Removal of IPinfo sources as input} \\ %
	    2024-11-07 & \textbf{Runtime optimizations} \\ %
        \bottomrule
    \end{tabular}
\end{table}

\begin{figure*}
    \centering
    \includegraphics[width=\linewidth]{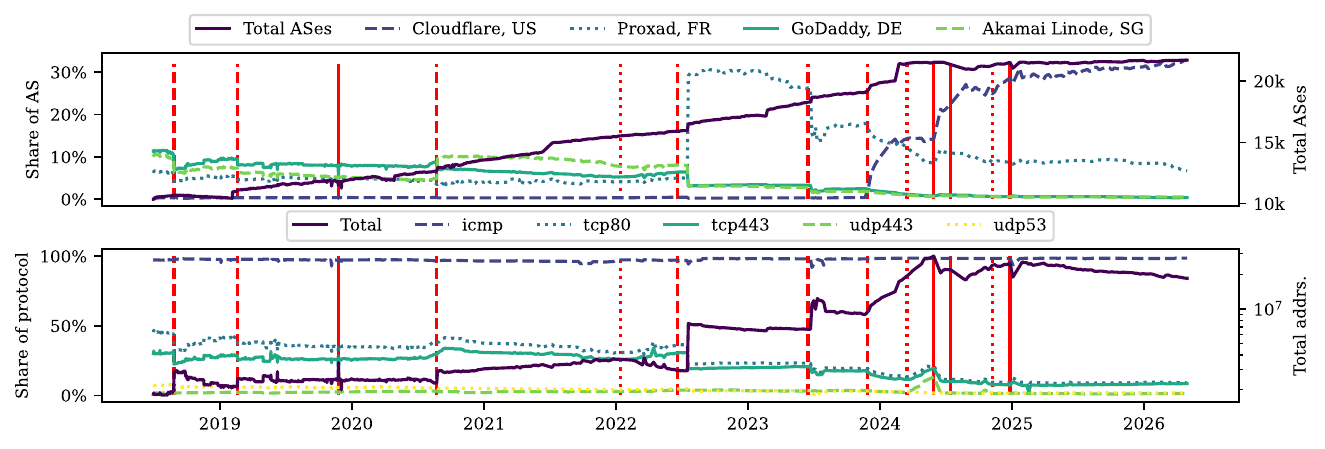}
    \caption{Development of the \hl measured by different metrics. Vertical lines represent the addition of selected new sources (dashed), major changes to the pipeline (dotted), changes without a corresponding change to the service (solid).}
    \label{fig:as-over-time}
\end{figure*}

\begin{figure}
    \centering
    \includegraphics[width=\linewidth]{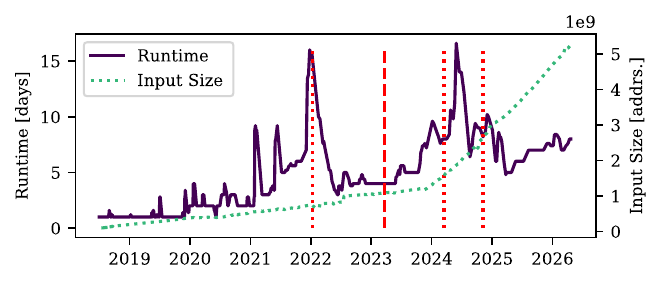}
    \caption{Temporal development of the scan runtime of the \hl compared to the total input size. Dashed vertical lines represent the addition of selected new sources, dotted vertical lines represent major changes to the pipeline.}
    \label{fig:scan-runtime}
\end{figure}

\parx{Technical Changes:}
Next, we summarize the most important technical changes made to the \hl pipeline over time; a summary can be found in \Cref{tab:changes}.
The service ran without major changes until 2021, when first scans with \ac{mptcp} probes were introduced, then UDP probes for port 161 with an SNMPv3 payload were added.
As part of the 2022 paper~\cite{zirngibl2022rustyclusters}, we identified many DNS responses as injections by the \ac{gfw} and filtered them.
This not only improved the quality of our scan results, but also greatly reduced the runtime of the subsequent scans as visible in \Cref{fig:scan-runtime}.
Since this filter still did not remove all injections from the \ac{gfw}, we changed the queried domain in the DNS probe from \texttt{www.google.com}, which is blocked by the \ac{gfw}, to a custom, non-blocked domain.
Between November 2024 and April 2025, we implemented optimizations which decreased the scanning runtime, see \Cref{fig:scan-runtime}.
Other upward spikes in scanning runtime are due to outages and hardware migrations.

\setlength{\belowcaptionskip}{-8pt}

\setlength{\belowcaptionskip}{0pt}

\begin{table}
    \centering
    \caption{Input sources of the \hl. Regular sources are included for each new scan. One shot sources are fixed sets of addresses being added only once or a few times. The first scan of the \hl was conducted and published on July 01, 2018, using address sources marked as \emph{First scan}. Some sources were excluded later for different reasons, e.g., data unavailability. Bold sources are highlighted in subsequent plots.}
    \label{tab:sources}
    \footnotesize
    \begin{tabularx}{\linewidth}{l p{3.5cm} X}
        \toprule
	    \textbf{Type} & \textbf{Timespan} & \textbf{Source} \\
        \midrule
	    Regular & First scan & DNS resolutions + AXFR \\ %
        Regular & First scan -- 2023-04-01 & Bitnode addresses \\ %
        Regular & First scan & Traceroutes from last scan \\ %
        Regular & First scan & RIPE Atlas traceroutes \\ %
        Regular & First scan & Open IP map \\ %
        Regular & First scan, 2020-08-24 -- 2022-05-31, 2023-02-24 -- 2025-04-16 & \textbf{Rapid7 DNS ANY} \\ %
        One shot & 2018-08-27 & \textbf{rDNS data (\citeauthor{fiebig2017rdns}~\cite{fiebig2017rdns})} \\ %
        One shot & 2019-02-19 & \textbf{rDNS data} \\ %
	    One shot & 2022-06-20 & \textbf{TGA addresses (IMC)} \\ %
	    One shot & 2023-06-16 & \textbf{TGA addresses (TMA)} \\ %
	    Regular & 2023-11-27 -- 2025-01-22 & \textbf{IPinfo 1} \\ %
	    Regular & 2023-11-27 -- 2024-03-16 & \textbf{IPinfo 2} \\ %
	    One shot & 2024-03-16 & 30-day-filtered addresses \\ %
        Regular & 2024-10-11 & CAIDA ITDK traceroutes \\ %
        \bottomrule
    \end{tabularx}
\end{table}

\parx{Changes to Input Sources:}
The address sources used in the \hl were subject to changes over time, as summarized in \Cref{tab:sources}.
While all sources added until November 2018 were already described in the 2018 publication by \citeauthor*{gasser2018clusters}~\cite{gasser2018clusters}, multiple new address sources were added afterwards.
We summarize these changes in the following and analyze their impact.
We mark changes with observable impact with dotted vertical lines in \Cref{fig:as-over-time}.
\citeauthor{fiebig2017rdns}~\cite{fiebig2017rdns} proposed a methodology to enumerate IPv6 addresses based on rDNS by exploiting the NXDOMAIN (denial of existence) semantics.
Data from their work was added to the \hl in 2018 and our own scans were conducted in 2019 following the same approach.
From mid 2020 onward, we included the Rapid7 DNS ANY dataset regularly, excluding it when our research access was not renewed, leading to an increase in total responsive addresses and a slightly increased share of addresses from AS63949 (Akamai Linode).
In 2022, following the publication by \citeauthor*{zirngibl2022rustyclusters}~\cite{zirngibl2022rustyclusters}, addresses generated by different \acp{tga} were included, resulting in a \sperc{173} increase in responsive addresses, a smaller increase in covered ASes and a shift in AS and protocol distribution.
The newly included addresses were mainly responsive to ICMP probes, with \sperc{27} from one AS (AS12322), which shifted the protocol and AS distributions.
In 2023, TGA outputs from another publication \cite{Steger2023} were included, inducing another increase in responsive addresses and covered ASes with a smaller share of the top ASes.
From November 2023 onward, two datasets from IPinfo~\cite{ipinfo} were regularly included.
This introduced a strong increase in IP addresses, covered addresses, but also scanning runtime, see \Cref{fig:scan-runtime} (dashed vertical line).
The latter was mainly caused by the increase in \texttt{/64} prefixes included in the cumulative input, each of which are scanned by our \ac{apd} mechanism.
This source also introduced a lot of addresses from AS13335 (Cloudflare), which shifted the share of represented ASes in the \hl.
The large number of addresses and other side effects led to the removal of these sources in 2024 and 2025 respectively, in turn leading to a (delayed) decrease in scan runtime.
Lastly, the source also led to an increase in total addresses in the \hl, while the number of addresses in aliased prefixes remained relatively stable, indicating the introduction of many non-aliased addresses.
An analysis of the APD results for prefixes from Cloudflare shows that we often receive fewer than 16 responses when evaluating subnets of aliased prefixes.
This is an anomaly \cite[Section 5.1]{gasser2018clusters}, which might be caused by ICMPv6 rate limiting.
\citeauthor*{Erdemir2025}~\cite{Erdemir2025} suggest adapting the response threshold from 16 to values as low as 10 to minimize this effect.
This shows that continuously adding sources adds value, but ongoing maintenance is required.

\parx{Automated Analysis of Changes:}
Lastly, we conduct an automated trend analysis of the data.
We start by comparing all scan dates that exhibit an absolute change in responsiveness of more than \sperc{120} compared to the next scan date.
We exclude dates which are already explained by changes in our sources or pipeline and analyze the remaining changes leaving us with the solid vertical lines in \Cref{fig:as-over-time}.
In late 2019, we observe a drop in responsiveness throughout all ASes and protocols, which was caused by errors in our pipeline following a hardware fault.
In mid 2024, we observe two strong decreases in responsiveness.
The IPinfo source introduced a constant number of active addresses, which however changed a lot between scan dates.
Addresses usually became unresponsive after few scans, being replaced by other responsive addresses, leading to a lot of address churn.
The removal of the IPinfo source led to the short-lived addresses not being replaced anymore, which in turn led to an overall decrease in responsiveness with substantial drops when bigger providers rotated addresses or prefixes.

\section{IPv6 Hitlist Coverage}
\label{sec:coverage}

While the \hl{} collects addresses from different sources, it is difficult to quantify its coverage of the IPv6 ecosystem.
To shed light on this question, we collaborate with a major central European ISP and Tier-1 provider and compare addresses of the \hl{} to visible traffic.
We use flow data collected at the edge of their network, aggregated per hour and /48 prefix of the source address.
The flow data is sampled and does not contain any information about flow targets.
Nevertheless, it allows us to provide an overview of the coverage of the \hl{} compared to real network traffic.

\begin{table}
    \caption{Coverage of ASes and /48 prefixes in the \hl{} (Apr. 30, 2026) compared to toplist resolutions in three months of flow data (containing at least one address or aliased prefix). Traffic indicates the share of IPv6 traffic these prefixes originate during the observation period.}
    \label{tab:flows}
    \footnotesize
    \begin{tabular}{llrrrrr}
        \toprule
        & & \multicolumn{2}{c}{ASes} & \multicolumn{2}{c}{/48} & Traffic [B] \\
        & & \# & \% & \# & \% & \% \\
        \midrule
        \multirow{8}{*}{\rotatebox[origin = r]{90}{\hl{}}} & Input & 21073 & 87.1 & 3899910 & 56.5 & 97.2 \\
        & Responsive & 18030 & 74.8 & 607863 & 8.8 & 54.7 \\
        & $\hookrightarrow$ ICMP & 17863 & 74.1 & 593607 & 8.6 & 53.3 \\
        & $\hookrightarrow$ TCP/80 & 10098 & 41.9 & 91068 & 1.3 & 52.4 \\
        & $\hookrightarrow$ TCP/443 & 9623 &  39.9 & 91551 & 1.3 & 52.7 \\
        & $\hookrightarrow$ UDP/53 & 7060 & 29.3 & 36554 & 0.5 & 11.7\\
        & $\hookrightarrow$ UDP/443 & 3358 & 13.9 & 12350 & 0.2 & 32.1 \\
        & Aliased & 677 & 2.8 & 32178 & 0.5 & 51.4 \\
        & Aliased + Responsive & 18169 & 75.4 & 637647 & 9.2 & 95.3\\
        \midrule
        \multirow{3}{*}{\rotatebox[origin = l]{90}{Toplists}}& 1-day  & 5432 & 22.5 & 24906 & 0.4 & 86.3  \\
        & 7-day & 5826 & 24.2 &  27718 & 0.4 & 88.5 \\
        & 30-day & 6233 & 25.9 & 30537 & 0.4 & 88.8  \\
        \bottomrule
    \end{tabular}
\end{table}

The dataset contains flows collected over three months (February -- April 2026).
It contains IPv6 flows from \sk{24.1} source ASes and \sm{6.9} /48 prefixes, with a daily median of \sk{13} observed ASes and \sm{1.3} /48 prefixes.
The total number of prefixes is influenced by a few ASes, \ie{} 10 ASes already account for \sperc{54.7} and 100 ASes account for \sperc{86.0} of the /48 prefixes, respectively.
A majority of these prefixes are only visible for 1--2 days, potentially indicating rotating prefix assignments in ISP networks.
\sk{58.9} further ASes are visible in the flow data.

We compare this flow data to addresses present in the \hl{} results from April 30, 2026 in three dimensions (ASes, prefixes, traffic covered) in \Cref{tab:flows}.
The cumulative input of the \hl{} contains addresses originated by \sk{21.1} ASes (\sperc{87.1}) and
\sm{3.9} of prefixes (\sperc{56.5}) visible in flow data.
Furthermore, the prefixes with at least one address known to the \hl{} account for \sperc{97.2} of the traffic volume captured in the available flow data.

Comparing only responsive addresses to our flow dataset shows that there is at least one responsive address in \sk{18} (\sperc{74.8}) ASes, mapped to only \sk{607.9} /48 prefixes (\sperc{8.8}).
Still these prefixes account for \sperc{54.7} of traffic captured by the flow data.
The coverage of protocol-specific responsive addresses follows an expected trend:
while addresses responsive to ICMP cover the most ASes and prefixes, addresses responsive to TCP/80 and TCP/443 cover prefixes resulting in slightly more traffic.
UDP/443 (the default port for HTTP/3 over QUIC) covers the fewest ASes and prefixes while still covering \sperc{32.1} of traffic.
As shown by \citeauthor{zirngibl2021over9000}~\cite{zirngibl2021over9000,zirngibl2023quic}, QUIC is mostly used by large CDNs.

We further map aliased prefixes identified by the \hl{} to visible /48 prefixes in the flow data.
Aliased prefixes are only identified in 677 ASes (\sperc{2.8}) with visible traffic and \sk{32.2} prefixes (\sperc{0.5}).
Nevertheless, these prefixes still cover \sperc{51.4} of traffic.
This aligns with findings by \citeauthor{zirngibl2022rustyclusters}~\cite{zirngibl2022rustyclusters}:
they showed that aliased prefixes are often seen in large CDNs and are most likely fully responsive prefixes used for load-balancing, and are not aliases of a single host.
Responsive and aliased prefix coverage can overlap due to the aggregation on /48 of flow traffic.
In combination, aliased prefixes and responsive addresses cover \sperc{95.3} of the traffic volume.
This reinforces their suggestion to include addresses (e.g., at least one) of aliased prefixes for IPv6 scans.

To show the usefulness of cumulative address collection over time, we compare the \hl coverage to the coverage of DNS resolutions of domain toplists, another technique commonly used for scans.
\Cref{tab:flows} shows the coverage of toplist-based lists of IPv6 addresses.
We use DNS resolutions of Google Crux~\cite{crux}, Cloudflare Radar~\cite{radar}, Majestic~\cite{majestic} and Cisco Umbrella~\cite{umbrella} from 1, 7, and 30 days (daily resolution of up-to-date lists respectively).
We do not test the responsiveness of addresses but evaluate the coverage based on IPv6 addresses from AAAA records.
Toplist addresses are mapped to \sk{30.5} /48 prefixes that still are the source of \sperc{88.8} of the traffic in bytes.
Many domains found in toplists are hosted by major CDNs or hypergiants accounting for large traffic shares.
However, compared to the \hl{}, fewer /48 prefixes and especially ASes are covered.
While toplists account for the top ASes, the long tail is easily missed by solely relying on DNS resolution of toplists.

In general, the evaluation shows good coverage of the \hl{} compared to traffic visible at a major central European ISP and Tier-1 provider.
At least one address is known to the \hl{} for \sperc{87.1} of ASes and \sperc{56.5} of /48 prefixes originating IPv6 traffic over a 3-month period and cover prefixes originating \sperc{97.2} of the IPv6 traffic.
\setlength{\belowcaptionskip}{-8pt}

\begin{figure}[tb]
  \centering
  \includegraphics[width=0.9\linewidth]{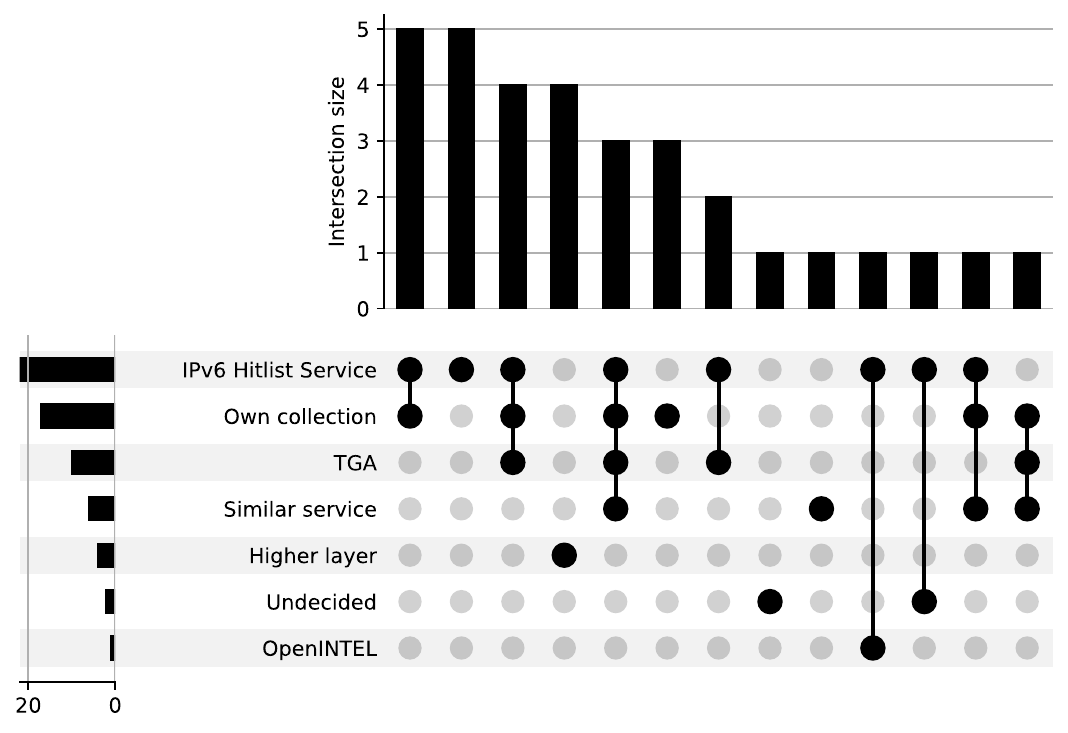}
  \caption{UpSet plot showing use of other address sources.}
  \label{fig:othersources}
\end{figure}

\setlength{\belowcaptionskip}{0pt}

\section{User Survey}
\label{sec:survey}

To better understand the current use of the \hl, we conduct a survey (see \Cref{app:survey} for details) and advertise it via networking mailing lists, Internet-measurement-related Slack and Mattermost channels, and on LinkedIn.
We also reach out to 147 registered \hl users via their registration emails.
In total, we receive 33 responses to our survey in a 30-day period (February 18, 2026 -- March 20, 2026).

\parx{IPv6 Measurements:}
We first ask survey participants, whether they are performing IPv6 measurements.
21 answered that they are currently performing IPv6 measurements, 8 did so in the past, and 3 plan to do so in the future.
Only one responded negatively to this introductory question.

\parx{Hitlist Use:}
We continue the remainder of the survey with the 32 responses which answered positively to the IPv6 measurement question.
Next, we inquire whether the respondents are using the \hl for their measurements.
22 responded that they are using the \hl, 9 responded that they are not using it, and one responded that they may be using it in the future.
Interestingly, we find that respondents who had performed IPv6 measurements in the past are more likely to make use of the \hl (\sperc{75}) compared to respondents currently running IPv6 measurements (\sperc{66}).
When asked about reasons of why they are not using the \hl, they indicated that they were not aware of it, or it did not align with their needs.

\parx{IPv6 Address Sources:}
Next, we analyze the use of other IPv6 address sources.
17 respondents use their own IPv6 address collection, 10 indicate the use of TGAs, and 6 responses indicate the use of a similar service as the \hl.
Four respondents indicate that they do not need dedicated IPv6 address targets as they are running their IPv6 measurements on a higher layer (e.g., domains, URLs).
Other respondents use OpenINTEL data~\cite{roland2016openintel}, are not using any additional address sources or still need to decide.
When looking at the combinations as shown in \Cref{fig:othersources}, we find that (1) using the hitlist or (2) using their own collection in combination with the hitlist are the two most common choices among respondents.
Furthermore, four respondents each are using their own collection with TGAs and the hitlist or do not need any dedicated IPv6 addresses as their measurements are run on a higher layer.
Some respondents also provide additional details on their IPv6 address target collection, stating that they collect addresses from various sources, \eg TLS certificates, toplists, TLD domain lists, /56 gateway addresses from LEO providers, running an NTP pool server, and traceroutes.

\parx{Hitlist Usage Patterns:}
To better grasp how the \hl is being used, we ask about usage patterns in the survey.
When it comes to duration, we find that \sperc{20} use the hitlist one-off, \sperc{15} use it for 1 month and another \sperc{15} for 2--6 months, \sperc{10} for 7--12 months, and \sperc{40} for more than 1 year.
Next, we ask all users about their frequency of use:
\sperc{5} use the service daily, \sperc{15} weekly, \sperc{25} monthly, \sperc{45} less frequently, and \sperc{10} as a one-off.
As the hitlist is available to users either publicly or through a registration-first service, we ask respondents how many are registered for the \hl.
16 respondents indicate that they are registered for the service, whereas 4 are not registered.

\parx{Hitlist Used Data:}
Next, we want to shed light on the \hl data used by researchers.
The service provides a publicly available list of responsive addresses and lists of aliased/non-aliased prefixes, which can be downloaded by everyone without registering an account (publicly available data).
Additionally, after registration researchers from academic institutions get access to additional data, \ie all input addresses (including non-responsive addresses), responsive addresses by different TCP, UDP, and ICMPv6 protocols, and access to historical data (registration-first data).
From the respondents, the majority (11) stated that they are using the publicly available as well as the registration-first data.
Moreover, 5 respondents use only publicly available data, whereas the remaining 4 respondents use only registration-first data.

Of the 15 respondents using the registration-first data, the majority use multiple types of data: 4 use responsive addresses, all addresses, and aliased data; 3 use responsive and aliased data; 3 use only responsive data; 2 use all addresses and aliased prefixes; another 2 use all addresses and responsive addresses; and a single respondent was using only the list of all addresses.
This underlines the heterogeneity of use cases for researchers using the service.

\parx{Measurement Types, Tools, and Aliasing:}
To better understand what types of IPv6 measurements researchers using the \hl perform, we next ask questions about targeted device types, used measurement tools, and handling of aliased prefixes.
Looking at the responses we find two equally important types of devices.
12 responses indicate mainly targeting servers (HTTP, DNS, etc.) and routers in their measurements.
Clients on the other hand are targeted by 9 respondents.
A few single respondents provided custom answers, highlighting that they are targeting exclusively DNS servers, a single address per /48 prefix, or all IPv6 hosts.
Note that respondents could provide multiple answers.

In terms of tooling, the majority of respondents use the port-scanning tool ZMapv6 \cite{zmapv6} or the application-layer scanning tool ZGrab2 \cite{zgrab2}, with 13 and 9 respondents, respectively.
Moreover, traceroute is used by 6 respondents, scamper \cite{luckie2010scamper} by 5 respondents, yarrp6 \cite{beverly2016yarrp} by 4 respondents, and Nmap \cite{lyon2009nmap} by 3 respondents.
Finally, DoE-Hunter \cite{doe-hunter}, ZDNS \cite{izhikevich2022zdns}, MAnycastR~\cite{10.1145/3730567.3764484}, and XMap \cite{li2026xmap} each saw a single mention among survey participants.
Note that respondents could provide multiple answers.

When asked about how survey respondents handle aliased prefixes in their measurements, the majority (8) rely on the provided lists by the \hl, while 4 respondents use those lists in addition to running their own alias prefix detection measurements.
6 respondents do not filter out aliased prefixes at all.

\parx{Impact:}
Next, we try to assess the academic impact of the \hl in terms of publications.
We ask survey participants whether their research using the \hl resulted in a research paper.
14 respondents indicated that this research resulted in research output in the form of a publication:
for 9 respondents this was the case for multiple publications, while for 5 respondents this resulted in a single publication.
For the remaining 6 respondents this has not resulted in a publication, of which 1 has a paper under submission and another 2 are planning a submission at the time of the survey.
We also ask survey participants about their perceived importance of the \hl to their own research, where 9 responded that it is very important, and they could not do their research without it; 11 participants responded that it is important as parts of their research rely on the \hl.
No participant responded that the hitlist was not important.
When asked about whether the \hl should be continued, 19 out of 20 respondents answered with yes, 1 answered with neutral.

These responses---in addition to the 240 citations for the IMC 2018 IPv6 Hitlist paper \cite{gasser2018clusters} as of May 2026---further underline the importance and impact that the \hl has on IPv6 measurement research.

\parx{Other Feedback:}
Finally, we ask survey participants for additional feedback to better understand \hl user needs.
We receive several custom responses, with ideas such as providing an API to retrieve the data, categorizing IP addresses by responsiveness over time or likelihood of being responsive in scans, and providing subsets of the hitlist for router addresses, servers, CDNs, etc.
We also receive several thank-you notes in this free-form feedback.

\section{IPv6 Hitlist Use}
\label{sec:hitlistuse}

Besides self-reported user behavior based on the survey, we also evaluate access patterns to the published data.
Therefore, we collect access logs of the webserver used to provide the data for 120 days.
We delete the logs after this study.
During this period, the \hl{} receives more than \sk{91} successful HTTP requests from 288 ASes (this might include crawlers and bots).
Next, we break those requests down into registered users and access to the openly available data.

Over the period of data collection 17 users and 2 measurement services~\cite{aschenbrenner2021from,albakour2021snmp} access results available via the registration-first service.
Registered users access different file types indicating different use cases and potential follow-up scans.
4 users access all three types of files, 4 users access the output and aliased files while 1 user accesses the input and aliased files.
The remaining 10 users focus on a single file type (3$\times$ input, 7$\times$ output).
Thus, the response files are the most commonly accessed files (15$\times$) followed by the aliased prefixes ($9\times$) and input ($8\times$).
As part of the \hl{}, continuous scans test responsiveness over different protocols (see \Cref{sec:technical}).
For those responsive IP addresses for different protocols we see different access patterns, \eg{} all protocols or only UDP/53 indicating interest in DNS specifically.
Besides different interest in available results, access patterns over time differ.
2 users access the data on one day only, 11 users on 2 to 9 days over multiple weeks.
4 users and the 2 measurement services access the data once per week.
One user downloaded the complete historical aliased prefix data going back all the way to 2018 within 3 days.
This shows that the availability of historic data can be useful for different purposes.
Furthermore, the available data granularity is actively used and indicates that the \hl{} data is used in a focused and careful manner, \eg{} to reduce the scale of follow-up scans. 

Next, we analyze access patterns of the publicly available \hl{} data.
As there is no user or registration information available for this data, we evaluate access based on an AS level in combination with user-agent information.
Throughout the 120-day period, we see 653 distinct AS-user-agent pairs accessing the publicly available data.
We discard individual IP addresses before conducting any analysis and do not store logs outside the scope of this study.
While this does not necessarily allow us to attribute individual Internet measurement or research entities, \eg{} if multiple users access the \hl{} from a cloud provider with the same user-agent, we are still able to analyze access patterns to the provided data.
Similar to registered users, access to publicly available data shows different patterns.
A majority of entities access the data on 5 days or fewer (609, \sperc{93}).
In contrast, 8 entities access the data daily throughout the 120-day period, \eg{} Shadowserver~\cite{Shadowserver}, Leitwert~\cite{leitwert}, or nxthdr~\cite{nxthdr}, most likely seeding regular follow-up IPv6 scans.
Besides registered academic users, measurement organizations frequently access the hitlist data over time.
\section{Best Practices \& Lessons Learned}
\label{sec:bestpractices}

\balance

\parx{Response Types:}
The \hl{} focuses on addresses with services resulting in positive responses, \ie{} sending an ICMP Echo Reply, a TCP SYN/ACK or a valid answer via UDP.
Thus, the lists of responsive addresses contain few routers that mostly send ICMP error messages or respond to specific services only, and few clients.
Furthermore, due to scalability reasons, the \hl{} focuses on specific ports and protocols (DNS and web).
While the list of responsive addresses to ICMP probes contains hosts potentially hosting different services, follow-up scans for services should consider an initial scan with the cumulative input. 

\parx{Cumulative Input:}
The input of the \hl{} is a combination of all addresses collected by the service since 2018 independent of whether they have been responsive to the tested protocols once or not.
While larger than the list of responsive addresses by multiple orders of magnitude, it is a valuable source for different goals.
For example, it contains router addresses extracted from traceroutes conducted by RIPE Atlas~\cite{ripe-atlas}, CAIDA Ark~\cite{caida-ark} and hitlist-specific yarrp scans.
Even if they do not return positive responses during follow-up scans, the input conserves these historical lists and can be used for future scans.
Furthermore, it might reveal changing address patterns or prefix rotation behavior of different networks.
TGAs could improve their evaluation of address patterns in networks by considering not only currently responsive addresses but also historical patterns.

\parx{Aliased Prefixes:}
Aliased prefixes are an important aspect to consider when running IPv6 measurements.
Often, they are completely filtered or ignored.
However, both extremes impact results.
On the one hand, ignoring aliased prefixes can easily skew address-based results or render scans infeasible.
Our provided list of aliased prefixes is based on the addresses known to the \hl{}.
Research adding other sources or using provided addresses, \eg{} for TGAs, needs to check for additional aliased prefixes to correctly interpret results.
On the other hand, removing aliased prefixes completely removes important parts of the Internet.
Aliased prefixes correlate with /48 prefixes originating \sperc{51.4} of traffic as seen in flow data (see \Cref{sec:coverage}).
As discussed by \citeauthor{zirngibl2022rustyclusters}~\cite{zirngibl2022rustyclusters}, aliased prefixes are not necessarily aliases used by a single host but often are fully responsive due to load-balancing artifacts from large CDNs.
Addresses from these prefixes should be sampled and used for follow-up scans to allow a representative view of the Internet.

\parx{Providing a Measurement Service:}
Over the past 10 years, we have learned important lessons on running a measurement service.
The initial version of the IPv6 Hitlist Collection, which ran from 2016 to 2018, only provided data to researchers upon request.
When relaunching the service as the \hl in 2018, we changed this model to a publicly available subset of the data, with raw data available after registration.
This flexible model allows everyone (researchers, network operators, even companies) to quickly download and work with the data, while it still allows researchers access to raw data if needed.
Furthermore, we can use registration emails to reach out to users of the service to announce new measurement data, published papers, or even conduct surveys, as we have done for this paper.
In general, we encourage researchers not only to publish data as a one-off dump for reproducibility purposes, but rather to consider running their measurements as a service to provide continuously updated data to fellow researchers.
Even though the work needed to maintain a measurement service is non-negligible, as things constantly change on the Internet, this changing nature also means that the results from long-running measurements often contain interesting insights which remain hidden when only performing one-off measurements.
Recently, we see a positive trend of more and more Internet measurement researchers adopting the model of running a measurement service accompanying their published paper~\cite{aschenbrenner2021from,albakour2021snmp,sattler2023hrp,izhikevich2024trust,fontugne2024yellow,khnsw-siius-25,hendriks2025laces,aleph2025}.

\section*{Acknowledgments}
This work was partially funded by the Horizon Europe programme under projects GreenDIGIT (101131207) and SLICES-IP (101287692), by the German Research Foundation (DFG) under project SLICES-SUSTAINABILITY (566292327) and by the German Federal Ministry of Research, Technology and Space (BMFTR) under projects ALL\#HANDS (02K25A026) and 6G-life (16KIS2414).

\bibliographystyle{ACM-Reference-Format}
\bibliography{paper,rfc}

\appendix
\section{Hitlist Pipeline}
\label{app:pipeline}

\begin{figure}
    \centering
    \footnotesize
    \begin{tikzpicture}[
        entity/.style={draw, minimum width=4cm, minimum height=1cm},
        innerentity/.style={draw, rounded corners=6pt, minimum width=3cm, minimum height={height("C") + 3mm}},
        inputentity/.style={draw, minimum width=2cm, minimum height={height("C") + 3mm}},
        scanentity/.style={draw, minimum width=4.5cm, minimum height=20pt},
        node distance=0.4cm
        ]

	    \node[innerentity] (input) {Combined Input Sources};
	    \node[inputentity, below=-.41pt of input,align=left] (inputs) {Cumulative input\\DNS Resolutions\\RIPE Atlas\\\ldots};

	    \node[innerentity, below=of inputs] (blocklist) {Blocklist Filter};
	    \node[innerentity, below=of blocklist] (apd) {Aliased Prefix Detection};
	    \node[innerentity, below=of apd] (unresponsive) {30-day Filter};

	    \node[scanentity, right=25pt of inputs] (traceroute) {Traceroute};
        \node[scanentity, below=5pt of traceroute] (scan) {ICMP TCP/80 TCP/443 UDP/53 UDP/443};
        \node[scanentity, below=5pt of scan] (scan2) {MPTCP/80 MPTCP/443 UDP/161};
	    
        \node[innerentity, below=of scan2] (responsive) {Responsive Addresses};

        \coordinate[below=5pt of unresponsive] (anchor1) {};
        \coordinate[right=1.7cm of anchor1] (anchor2) {};
        \coordinate[above=8pt of traceroute] (anchor4) {};
        \coordinate (anchor3) at (anchor4 -| anchor2);

        \draw[-latex] (inputs.south) -- node [midway,right] {5.24B} (blocklist.north); %
        \draw[-latex] (blocklist.south) -- node [midway,right] {5.16B (-81M)} (apd.north); %
        \draw[-latex] (apd.south) -- node [midway,right] {4.48B (-678M)} (unresponsive.north); %
        \draw[-latex, rounded corners] (unresponsive.south) to (anchor1)
            to (anchor2)
            to (anchor3)
            to node [midway,above] {458M (-4.02B)} (anchor4)
            to (traceroute.north); %
        \draw[-] (traceroute.south) -- (scan.north);
        \draw[-] (scan.south) -- (scan2.north);
        \draw[-latex] (scan2.south) -- node [midway,right] {19M} (responsive.north);
    \end{tikzpicture}%
    \caption{Pipeline of the \hl.}
    \label{fig:pipeline}
\end{figure}
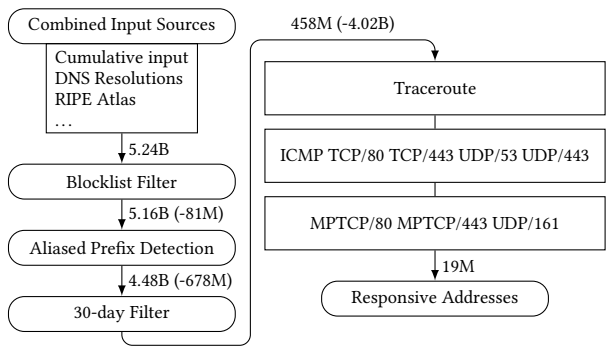

\Cref{fig:pipeline} provides an overview of the \hl{} pipeline collecting input sources, applying filters and conducting scans.
More details about the initial design can be found in~\cite{gasser2018clusters,zirngibl2022rustyclusters} and the current status and changes are discussed in \Cref{sec:technical}.

\section{Survey Details}
\label{app:survey}

Similar to related work in Internet measurement~\cite{xue2025towards,sultana2025survey}, we conduct a survey on IPv6 measurements and the use of the \hl.
We run the survey for a 30-day period (February 18, 2026 -- March 20, 2026) and receive 33 responses in total.
Below is the list of questions we asked in our survey.
Note that for some questions we had several variants (\eg different tense), depending on previous questions;
responses which are not present for all of these variants are denoted in parentheses.
Questions which allow multiple responses are prepended with ``Multi:'', optional questions are prepended with ``Optional:''.
Responses where participants could enter custom text are denoted as ``\textit{Other}'' or ``\textit{Text}''.

\subsection{Are you performing IPv6 measurements?}
\begin{itemize}
    \item Yes, currently
    \item Yes, I did so in the past
    \item Yes, I plan to do so in the future
    \item No
\end{itemize}

\subsection{Do/did you/do you plan to use the IPv6 Hitlist Service (ipv6hitlist.github.io) for your IPv6 measurements?}
\begin{itemize}
    \item Yes
    \item (Maybe)
    \item No
\end{itemize}

\subsection{Multi: Which other source(s) for IPv6 addresses do/did you/do you plan to use?}
\begin{itemize}
    \item Similar service providing a list of IPv6 target lists
    \item I have my own collection of IPv6 targets
    \item I'm using a seed set and then generating new addresses with a target generation algorithm (TGA)
    \item I don't need dedicated IPv6 targets as I'm running my IPv6 measurements on a higher layer (e.g., domains, URLs)
    \item (I still need to decide)
    \item None
    \item \textit{Other}
\end{itemize}

\subsection{Optional: Optionally provide more details on your exact IPv6 address source(s) and how you use/used/ plan to use them.}
\begin{itemize}
    \item \textit{Text}
\end{itemize}

\subsection{Optional: Optionally provide details on why you do not/did not/do not plan to use the IPv6 Hitlist Service for your IPv6 measurements.}
\begin{itemize}
    \item \textit{Text}
\end{itemize}

\subsection{Over which total duration do/did you download data from the IPv6 Hitlist Service?}
\begin{itemize}
    \item One-off
    \item 1 month
    \item 2--6 months
    \item 7--12 months
    \item More than 12 months
\end{itemize}

\subsection{How frequently do/did you download new data from the IPv6 Hitlist Service?}
\begin{itemize}
    \item Daily
    \item Weekly
    \item Monthly
    \item Less frequently than monthly
    \item One-off
\end{itemize}

\subsection{Did you register to use the IPv6 Hitlist Service data?}
\begin{itemize}
    \item Yes
    \item No
\end{itemize}

\subsection{Multi: Do/did you download publicly available data or data which is only available after registration for the IPv6 Hitlist Service?}
\begin{itemize}
    \item Publicly available data
    \item Data available after registration
\end{itemize}

\subsection{Multi: Which data do/did you download from the IPv6 Hitlist Service?}
\begin{itemize}
    \item All collected addresses (also called "input")
    \item Responsive addresses (also called "output")
    \item Aliased prefixes
    \item Non-aliased prefixes
\end{itemize}

\subsection{Multi: How do you filter aliased/highly-responsive prefixes from your IPv6 measurements?}
\begin{itemize}
    \item Using the list of aliased/non-aliased prefixes from the IPv6 Hitlist Service
    \item Running alias prefix detection measurements myself
    \item I do not filter them out
\end{itemize}

\subsection{Multi: Which tool(s) do/did you use to run your IPv6 measurements?}
\begin{itemize}
    \item ZMapv6
    \item ZGrab2
    \item yarrp6
    \item nmap
    \item traceroute
    \item scamper
    \item \textit{Other}
\end{itemize}

\subsection{Which types of devices are you targeting in your IPv6 measurements?}
\begin{itemize}
    \item Routers
    \item Servers (HTTP, DNS, etc.)
    \item Clients
    \item \textit{Other}
\end{itemize}

\subsection{Did the measurements where you used the IPv6 Hitlist Service result in a peer-reviewed publication?}
\begin{itemize}
    \item Yes, more than one
    \item Yes, one
    \item Not yet, it is currently under submission
    \item No, but I'm planning a submission
    \item No, no submission is planned at this point
\end{itemize}

\subsection{How important is/was data from the IPv6 Hitlist Service for your research?}
\begin{itemize}
    \item Very important, couldn't do it without it
    \item Important, parts of the research rely on it
    \item Not that important, I could do the research without the data
\end{itemize}

\subsection{Should we continue to take efforts to keep the IPv6 Hitlist Service running?}
\begin{itemize}
    \item Yes, the service is important for IPv6 measurement research
    \item Neutral, it doesn't hurt to have the data available
    \item No, the service is not needed anymore today
\end{itemize}

\subsection{Would you be willing to share IPv6 addresses/measurement data with the IPv6 Hitlist Service to further improve the service?}
\begin{itemize}
    \item Yes
    \item Maybe
    \item No
\end{itemize}

\subsection{Optional: Any other feedback, suggestions, improvements you want to share about the IPv6 Hitlist Service?}
\begin{itemize}
    \item \textit{Text}
\end{itemize}

\end{document}